\documentstyle[preprint,tighten,aps,psfig,epsf,floats]{revtex}

\newcommand{\be}{\begin{equation}}
\newcommand{\ee}{\end{equation}}
\newcommand{\ba}{\begin{eqnarray}}
\newcommand{\ea}{\end{eqnarray}}

\newcommand{\ep}{\epsilon}

\newcommand{\Dafne}{DA$\Phi$NE}

\begin{document}

\draft
\preprint{\vtop{\hbox{IEKP-KA/99-27}\hbox{SLAC-PUB-8322}
\hbox{RM3-TH/99-15}\hbox{hep-ph/0001064}
}}

\newcommand{\mar}{\marginpar{***}}

\title{ Contribution of the direct decay
$\phi\rightarrow\pi^+ \pi^- \gamma$  
to the process $e^+ e^- \to \pi^+ \pi^- \gamma$
at DA$\Phi$NE}

\author{Kirill Melnikov\thanks{
e-mail:  melnikov@slac.stanford.edu}}
\address{Stanford Linear Accelerator Center\\
Stanford University, Stanford, CA 94309}

\author{Federico Nguyen \thanks{
e-mail:NGUYEN@fis.uniroma3.it
}}
\address{
Dipartimento di Fisica dell'Universit\`a  and  INFN,   
sezione di Roma Tre, \\ 
Via della Vasca Navale 84, I-00146 Rome, Italy
}

\author{Barbara Valeriani \thanks{
e-mail:Barbara.Valeriani@pi.infn.it }}
\address
{Dipartimento di Fisica dell'Universit\`a and INFN, 
sezione di Pisa, \\
  Via Livornese 1291, I-56010 S. Piero a Grado (PI), Italy 
}

\author{Graziano Venanzoni,\thanks{
e-mail:Graziano.Venanzoni@iekp.fzk.de }}
\address{
Institut f\"{u}r Experimentelle Kernphysik,  
Universit\"{a}t Karlsruhe, \\
Postfach 3640, 76021 Karlsruhe, Germany}

\maketitle

\begin{abstract}
The potential of \Dafne\ to explore   direct radiative
decay $\phi \to \pi^+ \pi^- \gamma$ is studied in 
detail. Predictions of different theoretical 
models for this  decay are compared.
We find that it should be  possible to 
discriminate between these models at \Dafne\ in one year,
even assuming a relatively low luminosity 
${\cal{L}} = 10^{31}~{\rm cm}^{-2}\, {\rm sec^{-1}}$.
The influence of the decay $\phi \to \pi^+ \pi^- \gamma$ 
on the measurement  of total cross section $\sigma(e^+e^- \to {\rm hadrons})$
by tagging a photon in the reaction $e^+e^- \to \pi^+\pi^- \gamma$
is also discussed.
\end{abstract}

\section{Introduction}

Investigation of CP violation is the most important
physical goal of \Dafne, a high luminosity
$e^+e^-$ collider which operates on the $\phi(1020)$ 
resonance. However, thanks to high luminosity, 
there will be a substantial  amount  of data which may 
be used to advance our knowledge on
low energy hadron dynamics and even contribute to precision 
electroweak measurements \cite{j}. 

Recently it was suggested \cite{Binner}, that the  annihilation 
cross section $\sigma(e^+e^- \to  {\rm hadrons})$ 
at energies below the mass of the $\phi$ resonance may be 
studied at \Dafne\ using  reaction $e^+ e^- \to  {\rm hadrons} + \gamma$.
By tagging the photon it is possible  to determine 
the pion form factor at the  momentum transfer below 
the mass of the $\phi$ meson \cite{Binner}.  There are several possibilities
to improve further  the analysis of \cite{Binner} and in this 
paper we consider the contribution of the 
direct rare decay $\phi \to \pi^+ \pi^- \gamma$
to the reaction $e^+e^- \to \pi^+\pi^-\gamma$. 
Using the terminology of \cite{Binner},
the direct decay contributes to the final state radiation
which, for the purpose of the cross section measurement, has 
to be suppressed by an appropriate choice of  cuts on 
the photon and pion angles and energies. 
One of the aims of the present paper is to find out 
how the contribution 
of the direct decay affects 
 the analysis of Ref.\cite{Binner}. 

Besides that,  the rare decay 
$\phi \to \pi^+ \pi^- \gamma$ is  an interesting process by itself.  
As one deals here with the low energy limit of QCD, 
the first principles calculations are not possible and one has 
to resort to various models 
\cite{bcg,Close,LN,achas,oset,marco}. 
Since the number of models
is flourishing, we think that the experiments 
should distinguish  between them.
In principle, that can be achieved 
by studying the low energy region 
of the photon spectrum in the  reaction 
$e^+ e^- \to \pi^+ \pi^- \gamma$ \cite{JLF,novo},
but it is not an  easy task.
The reason is that the relative phases of the 
direct decay $\phi\to \pi^+ \pi^- \gamma$
and  the pure QED processes
(initial (ISR) and final (FSR) state radiation)
are not predicted by these models. 
As a consequence, the sign of the interference
term\footnote
{Let us note that for  symmetric cuts on the pions angles   
the initial state radiation 
does not  contribute to the interference term because of charge 
parity conservation.} appears to be to a large extent arbitrary. 
If one assumes that the interference is destructive,
the branching ratio of the direct decay becomes very  small,  
${\rm BR}(\phi \to \pi^+ \pi^- \gamma) \approx 4 \times 10^{-5}$.
Under such circumstances, a detailed study of the decay
$\phi \to \pi^+ \pi^- \gamma$ will be rather difficult,
requiring high statistics and a careful control over  efficiencies 
in order to discriminate between different models
by fitting  the photon spectrum. 

In this paper  we  report on the implementation
of the direct decay $\phi \to \pi^+ \pi^- \gamma$  into the 
Monte Carlo event generator for pure QED process 
$e^+e^- \to \pi^+\pi^-\gamma$ described in \cite{Binner}. 
Our implementation permits to 
choose between  different
models for the decay $\phi \to \pi^+ \pi^- \gamma$.
A clear advantage of having a Monte Carlo event generator 
for these studies is that it allows to keep  control over
efficiencies and resolution of the detector, fine tuning of
the parameters and  also provides 
for the possibility to generate realistic distributions where the
reaction $e^+ e^-\rightarrow\pi^+\pi^-\gamma$ is accompanied by
radiation of photons collinear to electrons and positrons~\cite{Binner}.

\section{The matrix element for  $\pi^+ \pi^- \gamma$ final state}

The matrix element for the direct decay 
$\phi \to \pi^+ \pi^- \gamma$ is parameterized as:
\be
{\cal M}_0 = -i e f_\phi (Q^2,Qq) {\ep_\phi}^\mu d_{\mu\alpha} 
{\ep}^\alpha,~~~~~~
d_{\mu\alpha} = (Qq)g_{\mu\alpha} - q_\mu Q_\alpha,
\label{decampl}
\ee
where $\ep_{\phi}$ and $\ep$ are  polarizations of the $\phi$ meson 
and the photon, respectively; 
$q$ is the momentum of the photon and $Q$ is the momentum of the $\phi$.
The function $f_\phi(Q^2,Qq)$ in Eq.(\ref{decampl}) is the 
form factor for the direct decay.
Its exact form depends on the chosen model.

Considering production of the $\phi$ meson in $e^+ e^-$ collision 
with the center of mass energy squared $s=Q^2$ and its 
subsequent decay to $\pi^+\pi^-\gamma$ final state, we find:
\be
{\cal M}_\phi = \frac{-ie^3}{Q^2} F_\phi (Q^2,Qq) \bar v(p_2) \gamma_\mu u(p_1)
d_{\mu\alpha}  {\ep}^\alpha,
\ee
where the form factor $F_\phi$ is defined as:
\be
F_\phi = \frac {g_{\phi \gamma} f_\phi}{Q^2 - M_\phi^2 + i M_\phi
\Gamma_\phi}.
\ee
The coupling constant $g_{\phi \gamma}$ describes
the mixing of the photon  and the $\phi$ meson
and can be determined from the decay
width of the $\phi$ meson into electron positron pair. 
Using
\be
\Gamma(\phi \to e^+ e^- ) = \frac {4\pi}{3} 
\frac {g_{\phi \gamma}^2 \alpha^2}{M_\phi^3}
\ee
and  
$\Gamma(\phi \to e^+ e^- ) = 1.3246 \cdot 10^{-6}~{\rm GeV}$,
$M_\phi= 1.0194~{\rm GeV}$, one obtains
$g_{\phi \gamma}=7.929 \cdot 10^{-2}~{\rm GeV}^2$.

Consider now the QED process 
$e^-(p_1) + e^+(p_2) \to \pi^+ (\pi_1) + \pi^-(\pi_2)+ \gamma(q)$. 
The initial state radiation amplitude reads:
\be
{\cal M}_{isr} = \frac {-ie^3F_\pi(Q_1^2)}{Q_1^2} 
\bar v(p_2) \left [
\frac {\gamma_\mu (\hat p_1 - \hat q) \gamma_\alpha}{-2p_1q}
+
\frac {\gamma_\alpha (-\hat p_2 + \hat q) \gamma_\mu}{-2p_2q}
\right ] u(p_1) \pi^\mu \ep^\alpha,
\ee
where $Q_1 = \pi_1 + \pi_2$ and $\pi = \pi_1 - \pi_2$.

For the amplitude of the final state radiation  we obtain:
\ba
&&{\cal M}_{fsr} = \frac {ie^3 F_\pi(Q^2)}{Q^2}
 \bar v(p_2) \gamma_\mu u(p_1)
\nonumber \\
&&\times
\left [ \frac {2\pi_2^\alpha + q^\alpha}{2\pi_2q}(-\pi^\mu +q^\mu)
+\frac {(-2\pi_1^\alpha-q^\alpha)}{2\pi_1 q} (-\pi^\mu - q^\mu)
 -2 g^{\alpha \mu}
\right] \ep_\alpha .
\ea

Then, the differential cross section for $e^+ e^- \to \pi^+ \pi^- \gamma$
can be written as:
\be
{\rm d}\sigma_{\rm total} = {\rm d}\sigma _{\rm QED} + {\rm d}\sigma _{\phi},
\label{total}
\ee
where ${\rm d}\sigma _{\rm QED}$ is the contribution considered in
\cite{Binner}
\be
{\rm d}\sigma _{\rm QED} \sim | {\cal M}_{isr} + {\cal M}_{fsr} | ^2,
\ee
\label{qed}
and 
\be
\label{e2}
{\rm d}\sigma _{\phi} \sim \left [ |{\cal M}_\phi|^2 + 2{\rm Re} \left \{
{\cal M}_{isr}^* {\cal M}_\phi \right \} 
+ 2{\rm Re} \left \{ {\cal M}_{fsr}^* {\cal M}_\phi \right \} 
\right ]
\ee
includes the amplitude of the direct decay. One sees that
${\rm d}\sigma_{\phi}$ contains different interference terms.
For this reason one might expect a significant dependence 
of the $\phi \to \pi^+ \pi^- \gamma$ signal 
on the relative phases of ${\cal M}_\phi$ and
${\cal M}_{isr}+{\cal M}_{fsr}$.
We will show below that this is indeed the case.
We now describe three different models for the
direct decay $\phi \to \pi^+ \pi^- \gamma$ that are 
implemented in our event generator.

\vspace*{0.3cm}
{\it  1. ``No structure'' model} \cite{bcg}.

In this case the decay $\phi \to \pi^+ \pi^- \gamma$
occurs through two subsequent 
transitions $\phi \to f_0 \gamma \to \pi^+ \pi^- \gamma$.
The form factor $f_\phi$ in Eq.(\ref{decampl})
becomes:
\be
f_\phi^{no\,str.} = \frac {g_{\phi f_0 \gamma} g_{f_0 \pi^+ \pi^-}}
 {m_{f_0}^2 -Q_1^2 -im_{f_0} \Gamma_{f_0}}.
\ee

The coupling constants in the above equation can be estimated
by using the information on the branching ratio 
${\rm Br}(\phi \to f_0 \gamma)$
 and on the branching ratio  
${\rm Br}(f_0 \to \pi^+ \pi^-)$. One obtains \cite{bcg}:
$$
|g_{\phi f_0 \gamma} g_{f_0 \pi^+ \pi^-}|
=(144 \pm 15) \sqrt{{\rm Br}(\phi \to f_0 \gamma)}.
$$

  Needless to say, that the region of applicability
of this model is restricted to relatively soft photons,
when the $f_0$ meson in the intermediate state 
is not too far off shell. For this 
reason, when implementing this model into the event generator,
we have introduced an additional exponential damping factor
which suppresses the emission rate for high energy photons \cite{JLF}:
\be
f_\phi^{no\,str.} \rightarrow 1.625 \times f_\phi^{no\,str.} 
 \exp{ \left \{ -\frac {(Qq)}{\Delta^2} \right \} },
\ee
with $\Delta=0.3~{\rm GeV}$ \cite{JLF}.

\vspace*{0.3cm}
{\em  2. $K^+K^-$ model} \cite{LN,Close}.

In this model one also has a two step transition, similar to 
``no structure'' model.
However, the $\phi \to f_0 \gamma$ decay amplitude 
is generated dynamically
through the loop of charged kaons.
The form factor $f_\phi$ in Eq.(\ref{decampl}) reads:
\be
f_\phi^{K^+ K^-} = 
\frac {g_{\phi K^+ K^-} g_{f_0 \pi^+ \pi^-} g_{f_0 K^+ K^-}}
{2\pi^2 m^2_{K} (m_{f_0}^2 -Q_1^2 -im_{f_0} \Gamma_{f_0})}
I\left ( \frac {m_\phi^2}{m_K^2},\frac {Q_1^2}{m_K^2} \right ).
\ee
The coupling constants $g_{\phi K^+ K^-}, g_{f_0 \pi^+ \pi^-}, 
g_{f_0 K^+ K^-}$ can be  estimated by using the information on corresponding
decay rates \cite{LN}:
\be
\frac{g^2_{\phi K^+ K^-}}{4\pi} = 1.66,
\,\, \frac{g^2_{f_0 \pi^+ \pi^-}}{4\pi m^2_{f_0}} = 0.105,
\,\, \frac{g^2_{f_0 K^+ K^-}}{4\pi} = 0.6~{\rm GeV}^2,
\ee
and $I(a,b)$ is the function known in the analytic
form \cite{LN,Close,Oller}:
\begin{equation}
\label{Iab}
I(a,b)=\frac{1}{2(a-b)}-\frac{2}{(a-b)^2}\left [
f(b^{-1})-f(a^{-1}) \right ]+
\frac{a}{(a-b)^2}\left [ g(b^{-1})-g(a^{-1})
\right ].
\end{equation}
The functions $f(x)$ and $g(x)$ are given by:
\begin{equation}
\label{fx}
\begin{array}{l}
f(x)=\left\{
\vspace{-.5cm}
\begin{array}{cl}
-\arcsin^2 \frac{1}{2 \sqrt{x}}~~~~~~~~~~~~~~~~~ x>\frac{1}{4},
\\ \frac{1}{4}[\ln \frac{\eta_+}{\eta_-}-i\pi]^2~~~~~~~~~~~~~~~~ x<\frac{1}{4},
\end{array}
\right.
\\[4.5ex]
 g(x)=\left\{
\vspace{-.5cm}
\begin{array}{c l}
\sqrt{4x-1} \arcsin \frac{1}{2 \sqrt{x}}~~~~~~~~~~x>\frac{1}{4},
\\ \frac{1}{2}\sqrt{1-4x} [\ln \frac{\eta_+}{\eta_-}-i\pi] ~~~~~~
~x<\frac{1}{4},
\end{array}
\right.
\end{array}
\end{equation}
with 
$ \eta_{\pm}= (1 \pm \sqrt{1-4x})/(2x)$.

\vspace*{0.3cm}
{\em 3. Chiral Unitary Approach ($U\chi PT$)} \cite{marco,Oller}.

In this case the decay $\phi \to \pi^+ \pi^- \gamma$
occurs through a loop of charged kaons that subsequently
annihilate into $\pi^+ \pi^- \gamma$.
The $f_0$ resonance is generated dynamically
by unitarizing the one-loop amplitude.
Using  notations of Ref.\cite{marco}, 
the form factor $f_\phi$ in Eq.(\ref{decampl})
reads:
\be
\label{eqm}
f_\phi^{U\chi PT} = t_{\rm ch} 
 \left\{\frac{G_V M_{\phi}}{f_\pi^2 2\sqrt{2}\pi^2 m_{K}^2}
I\left ( \frac {m_\phi^2}{m_K^2},\frac {Q_1^2}{m_K^2} \right )
+\frac{ \sqrt{2}}{M_{\phi} f_\pi^2}
\left(\frac{F_V}{2} -G_V\right) G_{K^+K^-} \right\},
\ee
where the coupling $G_V$ and $F_V$ are related to the decays
$\phi\to K^+ K^-$ and $\phi\to e^+ e^-$, respectively,
$f_\pi$ is the pion decay constant and $I(a,b)$ is the function 
given in Eq.(\ref{Iab}). $G_{K^+K^-}$ is defined by 
the integral:
\be
\label{intg}
G_{K^+K^-} = \frac {1}{2\pi^2}\int_0^{q_{\rm max}} \frac{q^2 dq}
{\sqrt{q^2+m_K^2} \left( Q_1^2 - 4(q^2+m_K^2) +i\epsilon \right )}.   
\ee
In Eq.(\ref{eqm})
$t_{\rm ch}$ is 
the strong scattering amplitude 
\begin{equation}
 t_{\rm ch} = \frac{1}{\sqrt{3}} t_{K\bar{K},\pi\pi}^{I=0}.
\end{equation}
The scattering amplitude $t_{K\bar{K},\pi\pi}^{I=0}$ 
is determined by using chiral perturbation theory (see Ref.\cite{oset}).
We have used the following values for the above constants:
$G_V= 0.055~{\rm GeV}$, $F_V= 0.165~{\rm GeV}$, $f_\pi=0.093~{\rm GeV}$,
$q_{\rm max}=0.9~{\rm GeV}$.

\begin{figure*}
\begin{center}
    \leavevmode
    \epsfxsize=14.cm
    \epsffile[19 22 638 744]{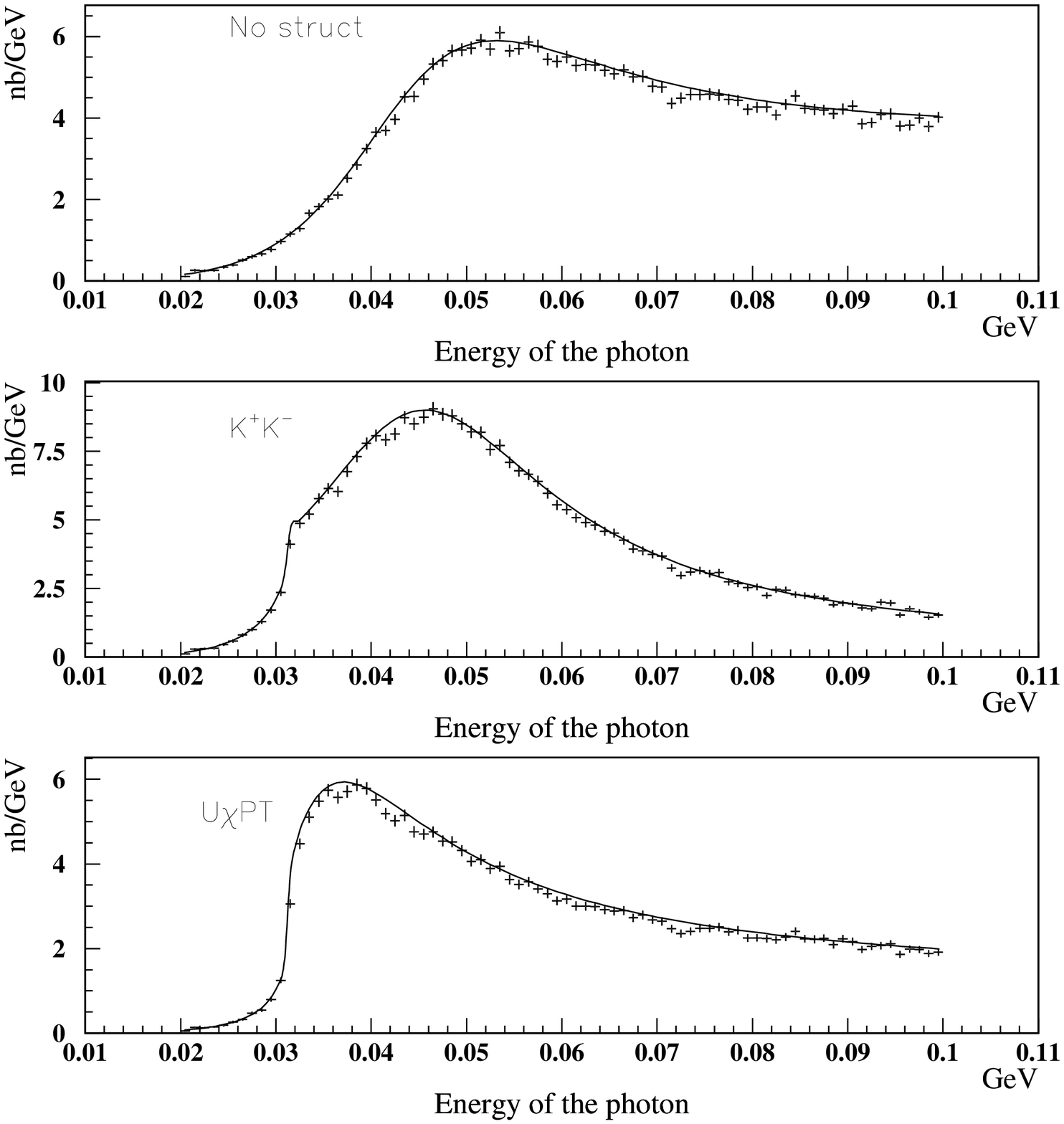}
    \hfill
    \parbox{14.cm}{
    \caption[]{\label{F1}\sloppy 
The energy  spectrum of photons produced in the direct decay 
$\phi \to \pi^+\pi^- \gamma$ without initial and final
state radiation. The results of the Monte Carlo 
simulation  are compared 
with the results of analytical calculations 
(solid curves) for different models \cite{bcg,LN,marco}. 
We use $m_{f_0}=0.976~{\rm GeV}$, $\Gamma_{f_0}=34~{\rm MeV}$.
For the ``no structure'' model the branching ratio
${\rm Br}(\phi \to f_0 \gamma)=10^{-4}$ has been used as an input
(this branching ratio is of the same order of magnitude as is predicted
by the other two models). For the purpose of  comparison, 
no exponential damping has been applied to the ``no structure''
model and the events were generated without collinear radiation.
We have applied the following cuts on the polar angle and the energy 
of the photons: $10^o < \theta_\gamma < 170^o$, $20~{\rm MeV}~< E_{\gamma}
 < 100~{\rm MeV}$.}}
  \end{center}
\end{figure*}

In Fig.\ref{F1} we present 
a comparison of the photon spectrum obtained using
the event generator and retaining only the term $|{\cal M}_\phi|^2$
in the cross section (cf. Eq.(\ref{e2})), with the 
analytic expressions from Refs.\cite{bcg,LN,marco}.
One sees a good agreement 
between the Monte Carlo simulation and the analytic results. 
Note also, that different models  predict  different 
shapes of the photon spectrum.


\section{Studying the direct decay $\phi\to\pi^+\pi^-\gamma$ at \Dafne }

We now address the question of whether 
precision studies of the 
direct decay $\phi\to\pi^+\pi^-\gamma$ 
are possible at \Dafne. 
While writing the general formula for the process
$e^+e^-  \to \pi^+\pi^-\gamma$, we have  
pointed out that the observable  signal of
$\phi \to \pi^+\pi^- \gamma$ 
might strongly depend on the interference with the FSR. The
$f_0$ signal may be enhanced if the sign 
between  $|{\cal M}_\phi|^2$ and 
$2{\rm Re} \left \{ {\cal M}_{fsr}^* {\cal M}_\phi \right \}$ is the
same (constructive interference) or may be reduced in the opposite case
(destructive interference).

\begin{figure*}
\begin{center}
    \leavevmode
    \epsfxsize=14.cm
    \epsffile[19 22 519 643]{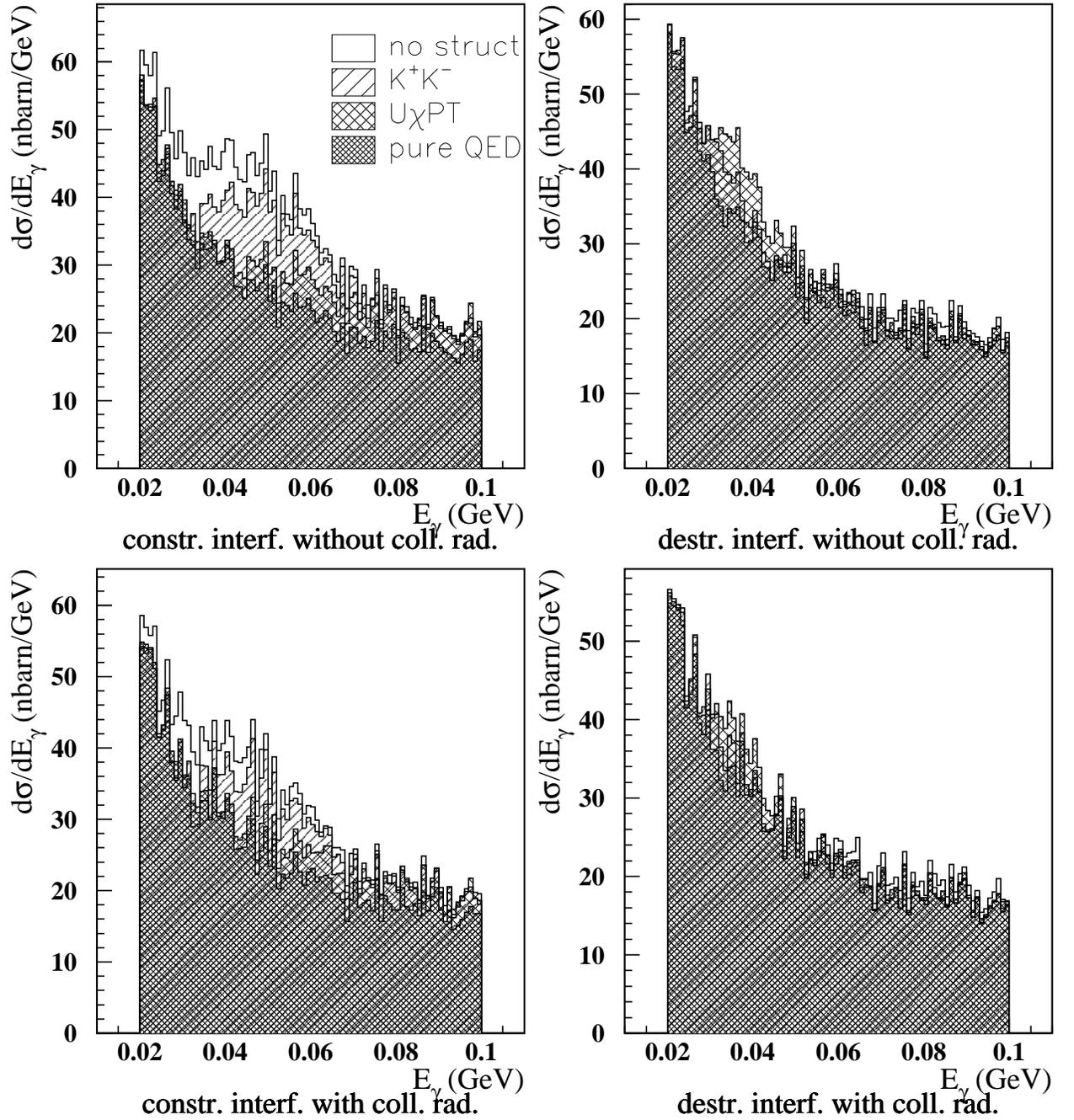}
    \hfill
    \parbox{16.cm}{
    \caption[]{\label{F3}\sloppy 
The spectrum of photons in the reaction
$e^+ e^- \to \pi^+ \pi^- \gamma$.
The following cuts were applied:
$|\cos{\theta_{\gamma}}|<0.9$, $|\cos{\theta_{\pi^{\pm}}}|<0.9$,
$|\cos{\theta_{\pi\gamma}}|<0.9$, where 
$\theta_{\gamma}$ and $\theta_{\pi}$ are  the polar angles of
photon and pions, respectively, 
and $\theta_{\pi\gamma}$ is the angle between the
photon and the $\pi^+$ momenta in the center of mass frame of 
the two pions. 
The exponential damping was applied to the  `no structure`'' model 
and the branching ratio  ${\rm Br}(\phi \to f_0 \gamma)=2.5\times 
10^{-4}$  has been chosen.
Different models are distinguished by different hatching. 
``Pure QED'' means that only 
the contribution due to  ${\rm d} \sigma_{\rm QED}$
is considered. 
}}
  \end{center}
\end{figure*}

In Fig.\ref{F3} we present the spectrum of photons in the reaction
$e^+ e^- \to \pi^+ \pi^- \gamma$ in the situation when the 
invariant mass of two pions is close to the mass of the $f_0$ meson.
We consider both constructive and destructive interference and  
generate events with and without collinear radiation, 
but initial and final state radiation is always kept.
One sees  from Fig.\ref{F3} that the collinear
radiation  results in the reduction of the signal.
However, if the tagged photon is emitted at a relatively large angle, the 
effect of collinear radiation can be partially removed by 
combining the information on  the position of the 
neutral cluster  in the calorimeter with the directions of the charged pions 
determined with the drift chamber. In this case the kinematics of the reaction 
becomes over-constrained and it is possible to restore the ``actual''
center of mass energy for any given event. This will require
a dedicated analysis, however.

\begin{figure*}
\begin{center}
    \leavevmode
    \epsfxsize=14.cm
    \epsffile[19 22 519 643]{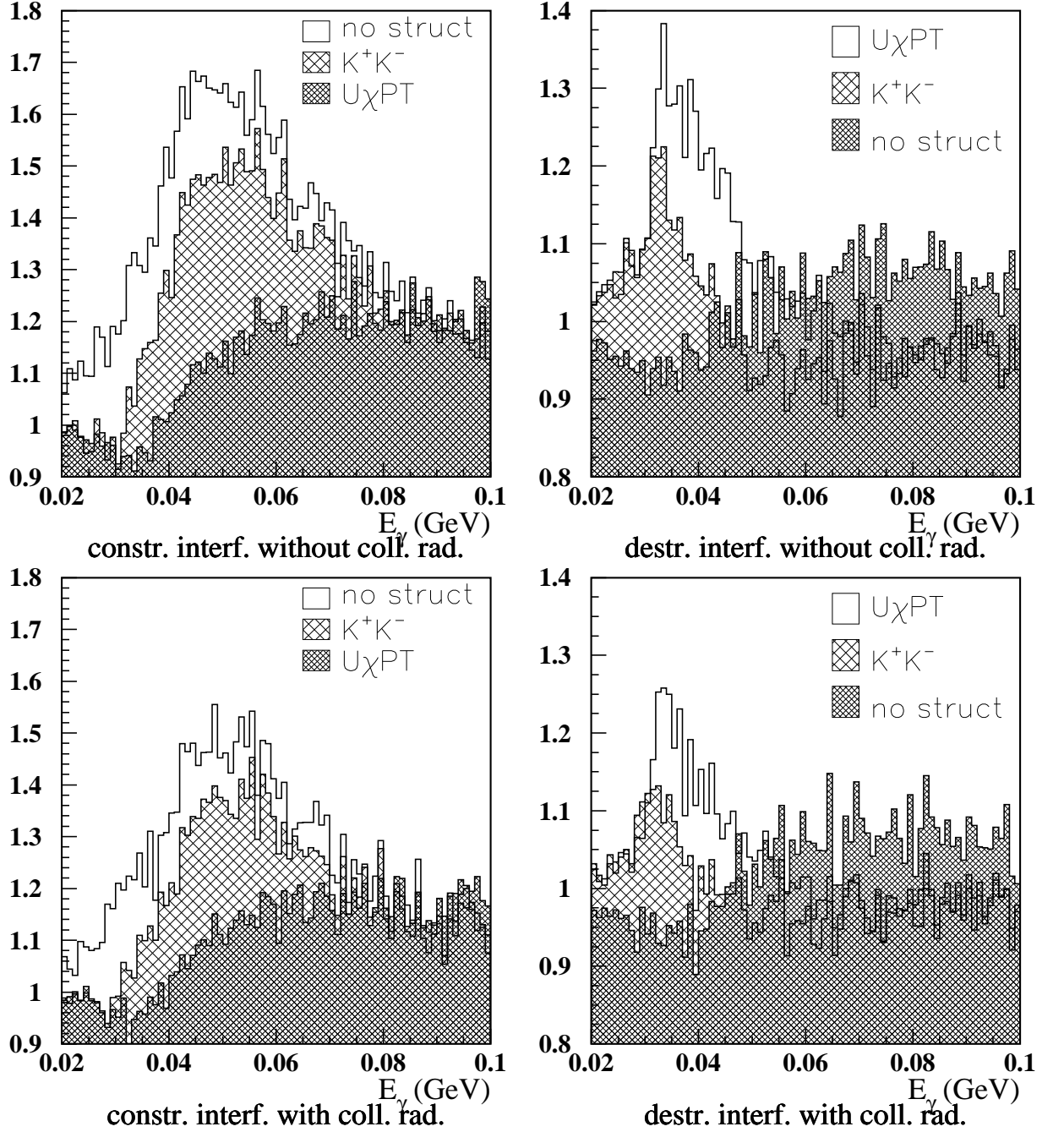}
    \hfill
    \parbox{14.cm}{
    \caption[]{\label{F4}\sloppy 
The signal-to-background ratio 
\( ({\rm d}\sigma_{\rm total}/{\rm d}E_{\gamma})
/({\rm d}\sigma_{\rm QED}/{\rm d}E_{\gamma}) \)
as a function of  photon energy.
The cuts are $|\cos{\theta_{\gamma}}|<0.9$,
$|\cos{\theta_{\pi^{\pm}}}|<0.9$,
$|\cos{\theta_{\pi\gamma}}|<0.9$. See text for more details.
}}
  \end{center}
\end{figure*}

Fig.\ref{F4} shows the signal-to-background ratio 
\be
S/B = \frac{{\rm d}\sigma_{\rm total}}{{\rm d}\sigma_{\rm QED}}
\ee
for different models.
As expected,  the sign of the interference affects not only the magnitude 
of the decay  $\phi \to\pi^+\pi^-\gamma$, 
but also the shape of the distribution.
The models where  the structure of the $f_0$ meson is assumed
show a broader signal for the constructive 
interference\footnote{The excess of events is significant in  
the region of photon
energies  $20~{\rm MeV} <E_{\gamma} < 100~{\rm  MeV}$.} 
than in the opposite case.
In addition,  the ``no structure'' model does not show a clear peak 
in the case of destructive interference and the $U\chi PT$ model in the case 
of constructive interference.

The number of events required to separate the signal 
from the background  can be obtained  by  estimating  the necessary 
number of events  in the energy region around the 
$f_0$ peak. We require the statistical error
to be smaller than  $10$\% of the signal itself. Hence,
\be
\delta N \le  \frac{\Delta N}{10},
\label{eq1}
\ee
where  $\Delta N$ is the number of  events due to  
direct decay of the $\phi$ meson:
\be
\Delta N = N_{\rm total} - N_{\rm QED}.
\ee
If we introduce a parameter $\xi$ such that
\be
N_{\rm total} = (1+\xi)N_{\rm QED},
\ee
Eq.(\ref {eq1}) takes the form:
\be
\delta N \le  \frac   {\xi  N_{QED}}{10}.
\ee
The value of  $\xi$ can be estimated from the $S/B$ ratio
shown in Fig.\ref{F4}.
Using the standard formula for 
statistical fluctuation, $\delta N = \sqrt{N_{\rm total}}$, we estimate
the number of events required to separate the contribution 
of $\phi \to \pi^+\pi^-\gamma$ from the QED background:
\be
N_{\rm QED} \geq  (1+\xi) \left( \frac{10}{\xi}  \right )^2,~~~~ 
{\cal{L}} \geq  \frac{N_{\rm QED}}{\epsilon \sigma_{\rm QED}}.
\label{24}
\ee
Here $\epsilon$ is the overall detector efficiency for  
$\pi^+ \pi^- \gamma$ events and ${\cal L}$ is the required 
integrated luminosity.

The results are summarized in
Table~\ref{tab1}. We use $\epsilon=0.5$ and 
 $\sigma_{\rm QED} = 2.1~{\rm nb}$ with the cuts  
$|\cos{\theta_{\gamma}}|<0.9$, $|\cos{\theta_{\pi^{\pm}}}|<0.9$,
$|\cos{\theta_{\pi\gamma}}|<0.9$ ( without collinear radiation).
When the  collinear radiation is included,  
 $\sigma_{\rm QED}$ is reduced to approximately $2~{\rm  nb}$.

\begin{table}
\begin{center}
\begin{tabular}{||c||c|c|c||c|c|c||} 
 \multicolumn{7}{||c||}{without collinear rad.}\\
\hline
Models &\multicolumn{3}{c||}{Construct. interf.} &  \multicolumn{3}{c||}
{Destruct. interf.} \\ 
\cline{2-7}
 & $\xi $ & $\# {\rm events}$ & ${\cal{L}} ({\rm pb}^{-1})$ 
&  $\xi $ & $\# {\rm events}$ & ${\cal{L}} ({\rm pb}^{-1})$ \\
\hline
No struct & 0.35 & 1100 & 1  
& 0.01  &$10^6$  & $\sim 2\times 10^3$   \\ 
~ & ~ & $20~{\rm MeV}<E_{\gamma}<100~{\rm MeV}$ 
& ~ &~ & $20~{\rm MeV}<E_{\gamma}<100~{\rm MeV}$ & 
 \\
\hline
$K^+K^-$ & 0.22 & $\sim 2100$  & $2$  & 0.07  & $22\times 10^3$ & 40  \\ 
~ & ~ & $20~{\rm MeV}<E_{\gamma}<100~{\rm MeV}$ 
& ~ &~ & $20~{\rm MeV}<E_{\gamma}<50~{\rm MeV}$ & 
 \\
\hline
$U\chi PT$ & 0.1 & $11\times 10^4$   & $\sim11$   & 0.15  & 5100  & 9 
\\ 
~ & ~ & $20~{\rm MeV}<E_{\gamma}<100~{\rm MeV}$ 
& ~ &~ & $20~{\rm MeV}<E_{\gamma}<50~{\rm MeV}$ & 
 \\
\hline
\hline
 \multicolumn{7}{||c||}{with collinear rad.} \\ \hline
Models &\multicolumn{3}{c||}{Construct. interf.} &  \multicolumn{3}{c||}
{Destruct. interf.} \\ 
\cline{2-7}
 & $\xi $ & $\# {\rm events}$ & ${\cal{L}} ({\rm pb}^{-1})$ 
&  $\xi $ & $\# {\rm events}$ & ${\cal{L}} ({\rm pb}^{-1})$ \\
\hline
No struct & 0.25 & $\sim 2000$  & $\sim 2$  & 0.01  & $10^6$  
& $\sim 2\times10^3$ \\ 
~ & ~ & $20~{\rm MeV}<E_{\gamma}<100~{\rm MeV}$ 
& ~ &~ & $20~{\rm MeV} < E_{\gamma}<100~{\rm MeV}$ & 
~ \\
\hline
$K^+K^-$ & 0.16 & $\sim 4500$  & $\sim 4$   & 0.04  &  $65\times 10^3$ &
118 \\ 
~ & ~ & $20~{\rm MeV}<E_{\gamma}<100~{\rm MeV}$ & ~ &~ & 
 $20~{\rm MeV} < E_{\gamma} < 50~{\rm MeV}$ & 
~ \\
\hline
$U\chi PT$ & 0.08 & $\sim 17000$   & 16  & 0.1  &  $11\times 10^3$  &
20  \\ 
~ & ~ & $20~{\rm MeV}<E_{\gamma}<100~{\rm MeV}$ 
 & ~ &~ & $20~{\rm MeV}<E_{\gamma}<50~{\rm MeV}$ & 
~ \\
\end{tabular}
\vspace*{0.5cm}
\caption{The number of events and the integrated luminosity 
required to observe  the direct decay
$\phi \to \pi^+ \pi^- \gamma$.}
\label{tab1}
\end{center}
\end{table}

In a similar way we make a rough estimate of  
the number of events and the luminosity
required to discriminate  between different models for the direct decay.
We obtain:
\be
N_{\rm QED}  \geq  \left( \frac{10}{\xi_{12}} \right )^2
, 
\ee
with
\be
\xi_{12} =  \xi_{1}-\xi_{2},
\ee
and $\xi_{1,2}$ are the $\xi$-parameters for  two models under consideration.
In Table~\ref{tab2}  we summarize the results.
One can see that the required number of events 
can be accumulated at \Dafne\ in less than one year 
assuming the  luminosity  ${\cal{L}} = 10^{31}~{\rm cm}^{-2}\, {\rm sec^{-1}}$.
The required luminosity is, however,  only indicative.

\begin{table}
\begin{center}
\begin{tabular}{||c||c|c|c||c|c|c||} 
 & \multicolumn{3}{c||}{without coll. rad.} &  \multicolumn{3}{c||}
{with coll. rad.} \\ 
\cline{2-7}
Models & \multicolumn{6}{c||}{Construct. interf.}    \\ 
\hline
 & $\xi_{12}$ & $\# {\rm events}$ & ${\cal{L}} ({\rm pb}^{-1})$ 
&  $\xi_{12}$ & $\# {\rm events}$ & ${\cal{L}} ({\rm pb}^{-1})$ \\
\cline{2-7}
No struct vs. $K^+K^-$ & 0.13 &    $6\times 10^3$ & $\sim 5$   & 0.09  & 
$\sim 11\times 10^3$ & $\sim 11$  \\ 
\hline
\hline
& \multicolumn{6}{c||}{Destruct. interf.}    \\ 
\cline{2-7}
 & $\xi_{12}$ & $\# {\rm events}$ & ${\cal{L}} 
({\rm pb}^{-1})$ 
&  $\xi_{12}$ & $\# {\rm events}$ & ${\cal{L}} ({\rm pb}^{-1})$ \\
\cline{2-7}
$K^+K^-$ vs. $U\chi PT$  & 0.08 &  $16\times 10^3$   & 29  & 0.06  & 
 $26\times 10^3$  & 47   \\ 
\end{tabular}
\vspace*{0.5cm}
\caption{The number of events and the integrated luminosity 
required  to distinguish  between two different models
for the rare decay $\phi \to \pi^+\pi^- \gamma$.} 
\label{tab2}
\end{center}
\end{table}

\section{Direct decay of the
$\phi$ meson and  the measurement of the electron positron
annihilation cross section at \Dafne}

\begin{figure*}
\begin{center}
    \leavevmode
    \epsfxsize=14.cm
    \epsffile[19 22 519 643]{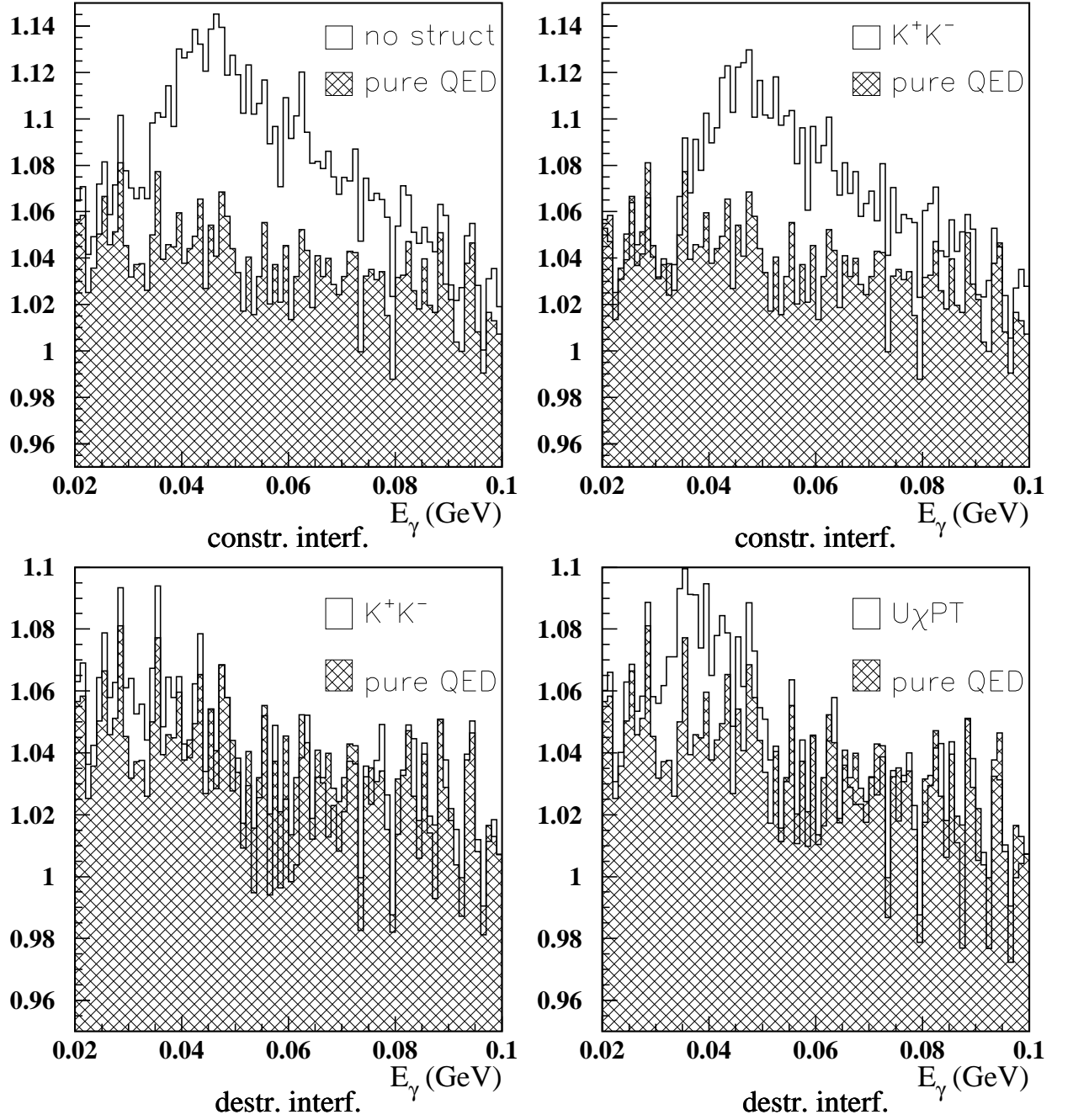}
    \hfill
    \parbox{14.cm}{
    \caption[]{\label{F5}\sloppy 
Ratio 
$({\rm d}\sigma_{{\rm total}}/{\rm d}E_{\gamma})
/({\rm d}\sigma_{\rm ISR}/{\rm d}E_{\gamma})$
as a function of the energy of the photon. The cuts are 
$7^o<\theta_{\gamma}<20^o$, 
$30^o<\theta_{\pi}<150^o$, and the invariant mass of  detected
particles in the final state $Q^2_{\pi^+\pi^-\gamma}> 0.9~{\rm GeV}^2$.
``Pure QED'' ratio is defined as
$({\rm d}\sigma_{\rm ISR+FSR}/{\rm d}E_{\gamma})
/({\rm d}\sigma_{\rm ISR}/{\rm d}E_{\gamma}.)$
Different pictures correspond to different models for the
$f_0$ resonance. The cases of constructive and destructive interference
are considered. See 
text for more details.}}
  \end{center}
\end{figure*}

It was suggested in Ref.\cite{Binner} that
the  measurement of  $\sigma (e^+ e^-\to {\rm hadrons})$
at \Dafne\ for different values of the center of mass energy 
can be performed by analyzing events with  additional hard photon emitted
at a relatively large angle ($\theta_{\gamma}>7^o$). 
The difficulties of this approach are related to the obvious
fact that  the hard photon 
can be emitted from both  initial and final state of the process.
If the ISR takes place, the total energy
of the collision is  reduced 
and such events can be used to measure 
$\sigma (e^+ e^-\to {\rm hadrons})$ at different energies.
In contrast to that, the photons caused by the FSR  represent
a background that must  be suppressed by applying suitable
cuts. Since, to be competitive\cite{j,Binner}, the measurement of 
$\sigma (e^+ e^-\to {\rm hadrons})$ for $\sqrt{s} < 1~{\rm GeV}$
has to  be performed at the one percent level, 
the practical realization of this idea is a non-trivial experimental task.
In Ref.\cite{Binner} only the QED process was studied.
Here we would like to add the direct decay $\phi \to\pi^+\pi^-\gamma$
which also contributes to the FSR and therefore increases the background. 
In Fig.\ref{F5} we show the values of
$({\rm d}\sigma_{{\rm total}}/{\rm d}E_{\gamma})
/({\rm d}\sigma_{\rm ISR}/{\rm d}E_{\gamma})$
with and without the contribution of the direct decay.
The ``pure QED'' case was studied in 
Ref.\cite{Binner}. The photon energies  
$20~{\rm  MeV} <E_{\gamma}<100~{\rm MeV}$ 
are considered, which  corresponds to  
$0.836~{\rm GeV}^2 <Q_{\pi^+\pi^-}^2<0.996~{\rm GeV}^2$. 
These invariant masses of two pions include
the contribution of the $f_0$ resonance and for this reason 
the largest contribution of the direct decay is expected in this
region. The cuts reduce the FSR considerably; nevertheless,
its contribution close to $f_0$ peak is significant.

\begin{figure*}
\begin{center}
    \leavevmode
    \epsfxsize=14.cm
    \epsffile[19 22  519 643]{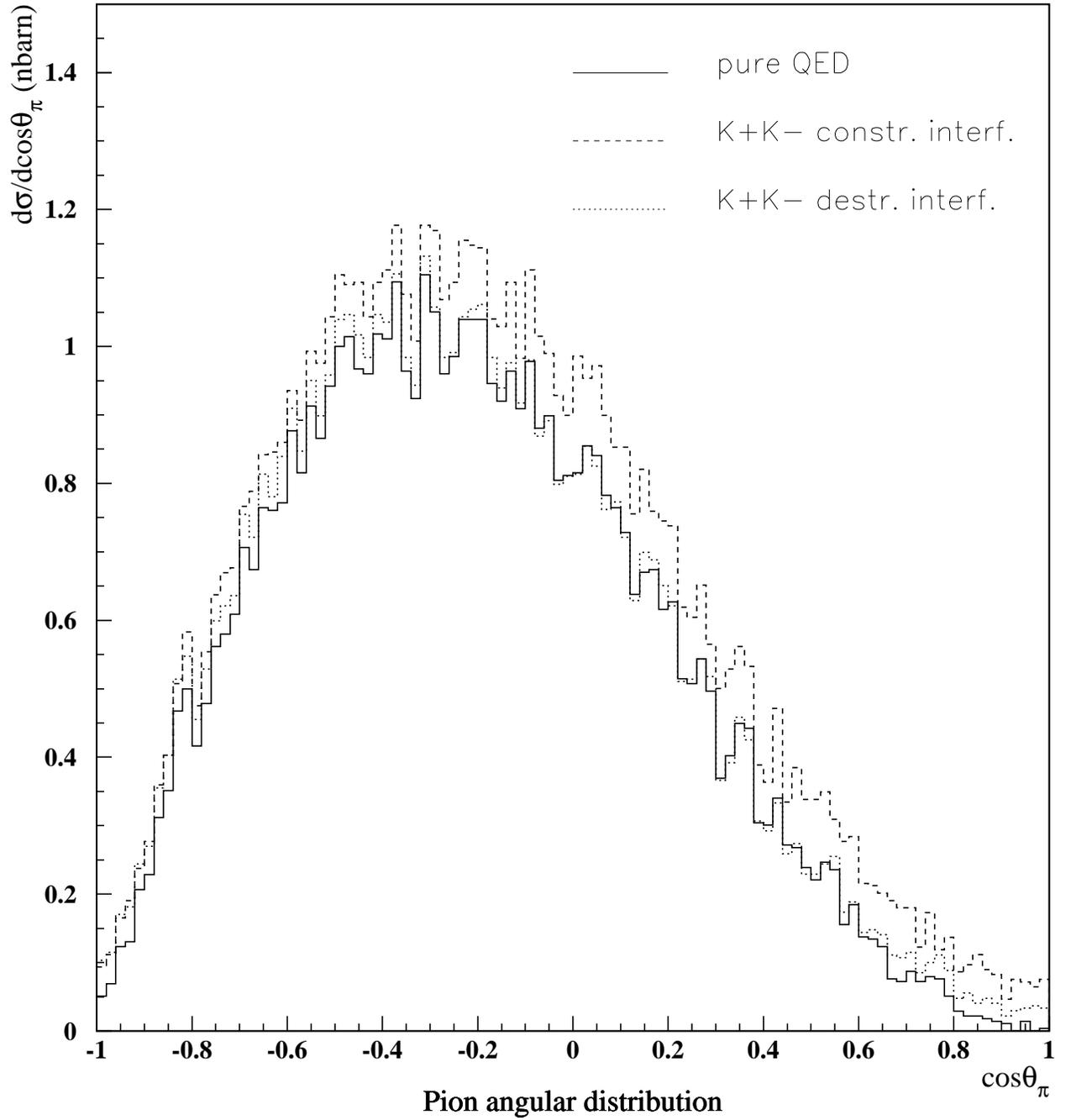}
    \hfill
    \parbox{14.cm}{
    \caption[]{\label{F6}\sloppy $\pi^+$ angular distribution. 
 The photon angle is $60^o<\theta_{\gamma}<120^o$.
The invariant mass of the two pions is 
$0.836~{\rm GeV}^2 <Q_{\pi^+\pi^-}^2<0.996~{\rm GeV}^2$.
}
}
  \end{center}
\end{figure*}

As discussed in \cite{Binner}, even ``pure QED'' 
theoretical predictions for the FSR are, strictly speaking, 
model dependent. It is therefore important to get a handle 
on it experimentally. In Ref. \cite{Binner} it was suggested to 
use the forward-backward asymmetry of the produced pions
to control the FSR.  The direct decay $\phi \to \pi^+ \pi^- \gamma$
changes the forward-backward asymmetry in the expected manner.
Since the contribution of the direct decay is significant only
if the invariant mass of the two pions is close to the mass of 
the $f_0$ meson, the forward-backward asymmetry
integrated over large range of $Q_{\pi^+\pi^-}^2 $
is not affected by the direct decay. Hence it can be used
to control the models for QED-like  final state radiation.
On the other hand, by  applying the cut
$0.836~{\rm GeV}^2 <Q_{\pi^+\pi^-}^2<0.996~{\rm GeV}^2$, we  significantly 
enhance the contribution of the direct decay to forward-backward
asymmetry. This is shown in Fig.\ref{F6} where 
predictions of $K^+K^-$ model are displayed
for both constructive and destructive interference.

\section{Conclusions}
We have discussed the contribution of the direct decay
$\phi \to \pi^+ \pi^- \gamma$ to the 
process $e^+ e^- \to \pi^+ \pi^- \gamma$
at \Dafne\ energies. To facilitate this study,  three
different models\footnote{
It is relatively straightforward to include other models
for the direct decay,
for example the
four quark model of Ref. \cite{achas}, to the event generator.
We plan to do that in the nearest future.}
for the 
direct decay $\phi \to \pi^+ \pi^- \gamma$ have been
implemented into the Monte Carlo event generator
described in Ref.\cite{Binner}.

The importance of this decay is twofold. First, it gives the  
information about the nature of the $f_0(980)$ meson.
Second, it provides an additional background to the measurement of 
$\sigma(e^+ e^- \to \pi^+ \pi^-)$ at different values of the center
of mass energy by tagging the hard photon in the reaction
$e^+ e^- \to \pi^+ \pi^- \gamma$.

We have  shown that \Dafne\ has a very good potential 
to study the nature of $f_0$ resonance. 
Even with   moderate luminosity 
${\cal{L}} = 10^{31}~{\rm cm}^{-2}\, {\rm sec^{-1}}$,
it is possible to discriminate between
different models for the decay $\phi \to \pi^+ \pi^- \gamma$
in a relatively short time.

As for  the measurement of the hadronic cross section 
$\sigma(e^+ e^- \to {\rm hadrons})$ at $\sqrt{s} < 1~{\rm GeV}$
using the process $e^+ e^- \to \pi^+ \pi^- \gamma$, 
we have found that the direct decay $\phi \to \pi^+ \pi^- \gamma$
increases the final state radiation by several 
percent in the region of pion invariant masses 
$0.836~{\rm GeV}^2 < Q_{\pi^+\pi^-}^2< 0.996~{\rm GeV}^2$, but quickly
dies out beyond this region. 

Finally, we note that it will also be possible to perform 
a detailed study of the decay $\phi\to \pi^o  \pi^o \gamma$
at \Dafne\ . We believe that  
it will be quite  useful to combine these
independent measurements in order  to check the theoretical
understanding  of the $f_0$ meson.

\section{Acknowledgments} 
We are grateful to J.H.~K\"{u}hn, W.~Kluge, G.~Pancheri, M.~Greco, A.~Denig
and G.~Cataldi for  useful discussions and
E.~Marco for providing us with his code. 
We  also thank M.Pennington for organizing a pleasant Euro\Dafne\
meeting in Durham  where part of this work was done.

This work is supported by the EU Network EURODAPHNE,
contract FMRX-CT98-0169, by BMBF under the contracts BMBF-06KA860
and BMBF-057KA92P, by the United States
Department of Energy, contract DE-AC03-76SF00515, 
by Gra\-duier\-ten\-kolleg ``Elementarteilchenphysik an Beschleunigern'' 
at the University of Karlsruhe and by the DFG Forschergruppe 
``Quantenfeldtheorie, Computeralgebra und Monte-Carlo-Simulation''


\end{document}